\documentclass[aps,preprint,prl,superscriptaddress,showpacs,amsmath,amssymb]{revtex4}

\usepackage{graphicx}
\usepackage{dcolumn}
\usepackage{bm}
\usepackage{color}

\newcommand{\lfp}{LiFePO$_4$}

\newcommand{\mbfu}{\(\mu _{\rm B}/{\rm f.u.}\)}
\newcommand{\tn}{$T_{\rm N}$}
\newcommand{\bsf}{$B_{\rm SF}$}

\begin{document}


\title{Antisite disorder in the battery material LiFePO$_4$}

\author{J.~Werner}
\email[Email:]{johannes.werner@kip.uni-heidelberg.de}
\affiliation{Kirchhoff Institute of Physics, Heidelberg University, INF 227, D-69120 Heidelberg, Germany}
\author{C.~Neef}
\affiliation{Kirchhoff Institute of Physics, Heidelberg University, INF 227, D-69120 Heidelberg, Germany}
\author{C.~Koo}
\affiliation{Kirchhoff Institute of Physics, Heidelberg University, INF 227, D-69120 Heidelberg, Germany}
\author{S.~Zvyagin}
\affiliation{Dresden High Magnetic Field Laboratory (HLD-EMFL), Helmholtz-Zentrum Dresden Rossendorf, D-01314 Dresden, Germany}
\author{A.~Ponomaryov}
\affiliation{Dresden High Magnetic Field Laboratory (HLD-EMFL), Helmholtz-Zentrum Dresden Rossendorf, D-01314 Dresden, Germany}
\author{R.~Klingeler}
\affiliation{Kirchhoff Institute of Physics, Heidelberg University, INF 227, D-69120 Heidelberg, Germany}
\affiliation{Centre for Advanced Materials (CAM), Heidelberg University, INF 225, D-69120 Heidelberg, Germany}

\date{\today}
\begin{abstract}

We report detailed magnetometry and high-frequency electron spin resonance (HF-ESR) measurements which allow detailed investigation on Li-Fe antisite disorder in single-crystalline \lfp , i.e., exchange of Fe$^{2+}$- and Li$^+$-ions. The data imply that magnetic moments of Fe$^{2+}$-ions at Li-positions do not participate in long-range antiferromagnetic order in \lfp\ but form quasi-free moments. Anisotropy axes of the magnetic moments at antisite defects are attached to the main crystallographic directions. The local character of these moments is confirmed by associated linear resonance branches detected by HF-ESR studies. Magnetic anisotropy shows up in significant zero-field splittings of $\Delta = 220(3)$~GHz, $\Delta'\sim50$~GHz and a highly anisotropic $g$-factor, i.e., $g_\mathrm{a} = 1.4$, $g_\mathrm{b} = 2.0$, and $g_\mathrm{c} = 6.3$. We demonstrate a general method to precisely determine Fe-antisite disorder in \lfp\ from magnetic studies which implies a density of paramagnetic Fe$^{2+}$-ions at Li-positions of 0.53\%.

\end{abstract}
\maketitle

\section{Introduction}
Electronic and ionic conductivity crucially affect the performance of electrode materials for lithium-ion batteries (LIB), including the charge and discharge rates, cycling stability and practically accessible capacity. Therefore, it is essential to understand the properties of defects unavoidably appearing in real materials as a precondition to understanding conductivity and eventually battery performance. In systems like the commercialised battery material \lfp, where ionic diffusion is supposed to be promoted along channels of the lattice structure, i.e., of one-dimensional (1D) nature, defects particularly affect actual conductivity as channels are easily blocked~\cite{gardiner2010anti}. However, defects are not necessarily detrimental to high-performance electrodes but may significantly enhance it by enabling additional low-energy ionic pathways~\cite{wang2007ionic,malik2010particle}. This is particularly true for Li-Fe anti-site disorder which is an intrinsic type of defect in olivine-like transition metal phosphates. Due to the similar covalent radii of Li- and Fe-ions both can change places and form a so-called anti-site defect. Calculations in Ref.~\onlinecite{yang2011li} suggest that ionic migration perpendicular to the 1D channels is energetically more favourable than along the blocked channels, and might even yield ionic migration of higher dimensionality by supporting channel crossover\cite{malik2010particle,gardiner2010anti}. In this respect, activation energies for ionic diffusion along the crystallographic directions perpendicular to the channels can be interpreted as activation of channel crossover via anti-site defects~\cite{liu2017effects}. It has been shown that the essentially 1D transport in \lfp\ becomes 3D when the percolation threshold of only a few \% of antisite disordered Li-/Fe-positions is reached~\cite{adams2010lithium} which evidently would have strong implication for understanding and designing actual battery materials. Ionic migration of higher dimensionality is indeed observed experimentally in single crystals studies~\cite{amin2008aluminium,neef2020anisotropic}. Understanding the nature of Li-Fe antisite defects is hence crucial to calculate relevant input parameters for modelling \lfp\ as an electrode material~\cite{dathar2011calculations}.

Experimental studies on antisite disorder are challenging as conventional x-ray scattering is not sensitive to Li. In this paper we apply magnetic probes to study the intrinsic properties of antisite defects in \lfp\ single crystals, i.e., tunable high magnetic field/high-frequency electron spin resonance (HF-ESR) spectroscopy and static magnetisation. We exploit the fact that antisite defects are magnetic in a sense that paramagnetic Fe-ions reside in different crystallographic environments and are hence distinguishable and detectable while, at low-temperatures below the magnetic ordering temperature, Li-ions form local non-magnetic defects in the long-range antiferromagnetic ordered magnetic lattice of Fe-moments. We show that the density of antisite defects is straightforwardly and precisely determined. In addition, energy levels of the Fe$^{2+}$-ions at the Li-position are studied in high magnetic fields by HF-ESR. We observe significant zero field splittings of $\Delta'\sim50$~GHz and $\Delta = 220(3)$~GHz of antisite Fe$^{2+}$-moments and highly anisotropic effective $g$-values along the different crystallographic axes, i.e., $g_\mathrm{a} = 1.4$, $g_\mathrm{b} = 2.0$, and $g_\mathrm{c} = 6.3$ for the $a$-, $b$-, and $c$-axis, respectively (space group $Pnma$). This anisotropy of the g-factor is resembled by magnetic susceptibility which shows a strongly anisotropic Curie-Weiss-like upturn at low temperatures.

\section{Experimental}

Single crystals of \lfp\ were grown by the high-pressure optical floating-zone method as reported in detail in Ref.~\onlinecite{neef2017high}. Magnetization in static magnetic fields up to 7~T was studied by means of a Quantum Design MPMS-3 SQUID magnetometer. HF-ESR measurements were performed in a transmission type ESR spectrometer in Faraday geometry. ESR measurements in static fields up to 16~T were done by means of a phase-sensitive millimeter-wave vector network analyzer (MVNA) from AB Millimetr\'{e}~\cite{comba2015magnetic}. HF-ESR experiments  in pulsed fields up to 50 T were performed at the Dresden High Magnetic Field Laboratory  (HLD) using VDI modular transmitters (product of Virginia Diodes Inc., USA) as sub-mm radiation sources and InSb hot-electron bolometer as a radiation detector.

\section{Results}

\begin{figure}[t]
\includegraphics[width=1.0\columnwidth,clip] {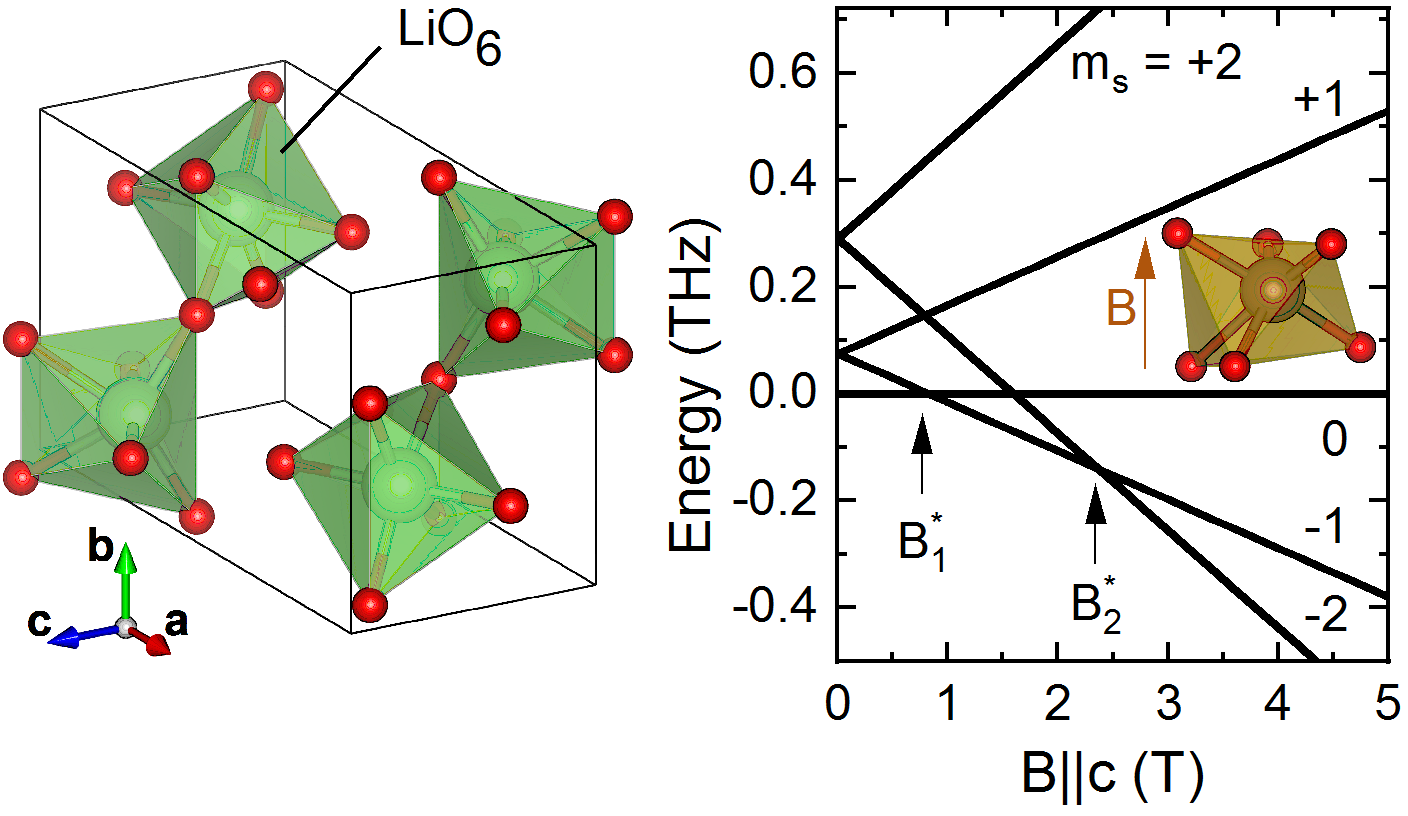}
\caption{(a) The four different LiO$_6$ octahedra of the \lfp\ unit cell. (b) Simplified energy-level diagram and crystal environment of the Fe-antisite defects with magnetic field applied along the crystallographic $c$-direction. Crystal structure visualizations were generated with VESTA \cite{momma2011vesta}. }\label{struc}
\end{figure}

\subsection{Magnetisation}

\begin{figure}[b]	
\includegraphics[width=1.0\columnwidth,clip] {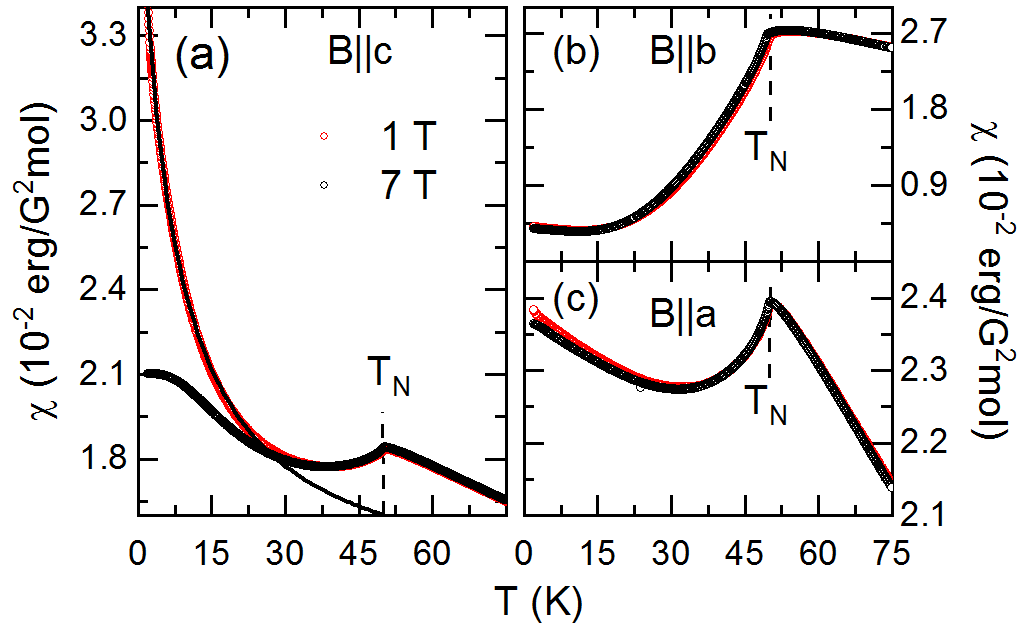}
\caption{Static magnetic susceptibility at $B=0.1$ and 7~T with external magnetic fields applied along (a) $B||c$ ,(b) $B||b$ and (c) $B||a$. The dashed lines indicate the N\'{e}el-temperature \tn$=50$~K of \lfp\ at 0.1~T. The solid black line in (a) is a Curie-Weiss fit of the data (see the text).}\label{MT}
\end{figure}

\lfp\ exhibits long range antiferromagnetic order of the $S=2$ spins of the magnetic Fe$^{2+}$-ions,
which evolves at \tn = 50.0(5)~K~ \cite{santoro1967antiferromagnetism,werner2019high,li2006spin,zhi2004mossbauer}. In the ordered phase, the spins are mainly directed along the crystallographic $b$-axis~\cite{rousse2003magnetic} with a small collinear rotation towards the $a$-axis and spin canting along the $c$-axis with an overall rotation of the ordered moments of 1.3(1)$^\circ$ off the $b$-axis~\cite{toft2015anomalous,yiu2017hybrid}. The onset of magnetic order, i.e., \tn , is reflected by kinks in the static magnetic susceptibility $\chi=M/B$ shown in Fig.~\ref{MT}, and the easy axis $B\|b$ is evident from Fig.~\ref{MT}b. Notably, in contrast to rather constant values of $\chi$ which are expected for a typical antiferromagnet~\cite{nagamiya1955antiferromagnetism}, the susceptibility $\chi(B=0.1~\rm{T}\|c)$, i.e., applied along the hard axis, strongly increases upon cooling (see Fig.~\ref{MT}a). We note that this magnetic field is far below any metamagnetic transition in \lfp~\cite{werner2019high}. In contrast, there is only a weak temperature dependence for $B\|a$-axis (intermediate axis).  Phenomenologically, the strong upturn of $\chi_{\rm c}$ is described by a Curie-Weiss law $\chi_{\rm b}=C_\mathrm{c}/(T+\Theta_\mathrm{c})$, with the Curie constant $C_\mathrm{c} = 0.179(3)$~erg K G$^{-2}$ mol$^{-1}$ and the Weiss-temperature $\Theta_\mathrm{c} = 6.5(1)$~K (see black line in Fig.~\ref{MT}a). This suggests the presence of only weakly correlated (i.e., 'quasi-free' with respect to the main antiferromagnetic ordering phenomenon at \tn ) magnetic moments which do not participate in long-range antiferromagnetic spin order. The fact that only a weak Curie-like increase is observed below 30~K for the other directions implies the anisotropic nature of these quasi-free spins~\cite{baumann2006anisotropic}. This is confirmed by application of $B=7$~T which nearly does not affect $\chi_\mathrm{a}$ and $\chi_\mathrm{b}$ while the steep upturn in $\chi_\mathrm{c}$ is suppressed towards a rather expected behaviour. \tn\ is almost unchanged either. Please note, that interpreting the steep upturn as the response of moments not involved in the overall spin structure implies that the parameters obtained by Curie-Weiss fitting must be taken cautiously as the response of the long range order phase is not precisely known. As will be shown, precise quantitative separation is feasible if $M(B)$ data are exploited.

\begin{figure}[htb]	
\includegraphics[width=1.0\columnwidth,clip] {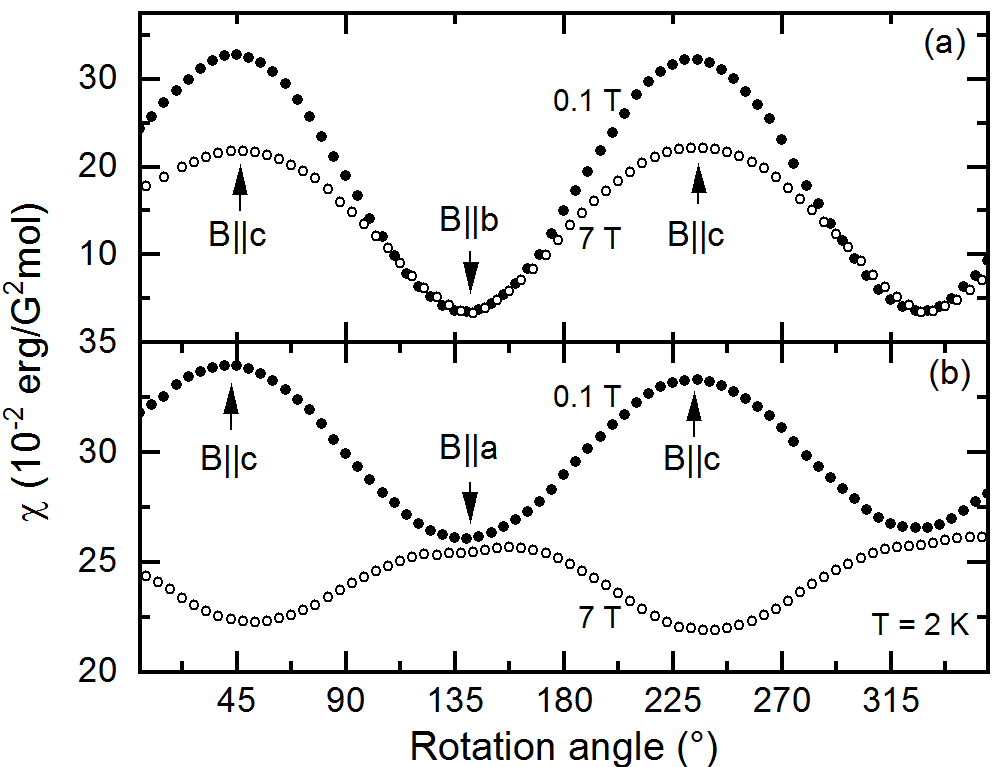}
\caption{Static magnetic susceptibility at $B=0.1$ and 7~T, at $T=2$~K, with external magnetic fields applied (a) in the $bc$-plane and (b) in the $ac$-plane. Arrows indicate the main crystallographic directions.}\label{rot}
\end{figure}

Anisotropy of the low-temperature magnetisation is further elucidated in Fig.~\ref{rot}, where the susceptibility at $T=2$~K is shown in dependence of the field direction for $B=0.1$ and 7~T. At $B=7$~T, the expected behaviour for antiferromagnets is observed, i.e., the susceptibility is smallest for $B||b$ (easy axis) while it is highest for $B||a$ (intermediate axis). This differs from the finding at small magnetic field where the quasi-free moments significantly contribute. At $B=0.1$~T, the susceptibility is almost unchanged with respect to the high-field values for $B||b$ and $B||a$. In contrast, $\chi_c$ is much larger which must be attributed to the quasi-free moments. The data hence imply that the $g$-tensor of the quasi-free moments is strongly anisotropic and attached to the main crystallographic directions of \lfp. The discrepancies of minima and maxima for the different measurements are within the error bars of the MPMS-3 rotor of approximately 5$^{\circ}$.

\begin{figure}	
\includegraphics[width=1.0\columnwidth,clip] {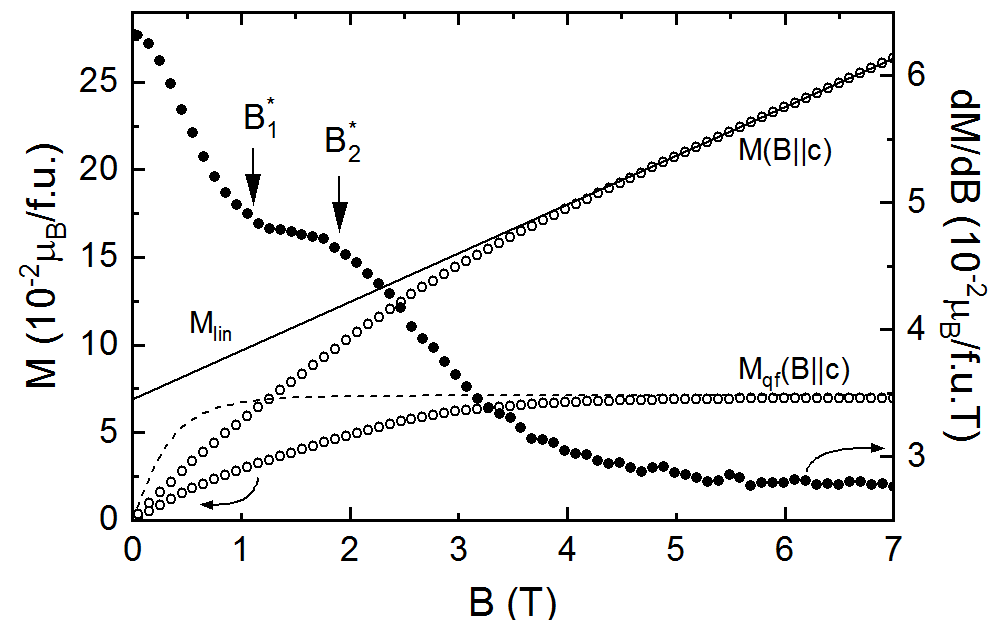}
\caption{Magnetisation (left axis) and its derivative $\partial M/\partial B$ (right axis), at $T=1.8$~K, for $B||c$. $M_\mathrm{qf}=M-M_{\rm lin}$ is the difference between the measured magnetisation $M$ and high-field linear contribution $M_{\rm lin}$ (see solid line), the dashed line represent a fit to $M_{\rm qf}$ by means of the Brillouin function and $B^*_1$, $B^*_1$ mark anomalies in $\partial M/\partial B$.}\label{MB}
\end{figure}

Measurements of the magnetization in dependence of the external magnetic field (Fig.~\ref{MB}) reflect the unusual anisotropic behaviour of the low-temperature susceptibility as well. While for magnetic fields $B\perp c$ only tiny right bending is observed, significant curvature is only observed for $B||c$ (cf. Ref.~\onlinecite{werner2019high}). This right bending is superimposed by a linear-in-field contribution to the magnetisation up to the highest measured fields of $\geq 56$~T~\cite{werner2019high}, except for metamagnetic transitions appearing at $B||b\sim 32$~T. Such linear field effect is typical for three-dimensional antiferromagnets and is attributed to the response of the antiferomagnetically ordered spins $S=2$. In contrast, right bending at rather low energy fields clearly signals the alignment of magnetic moments which are not expected in a conventional antiferromagnet. Note, that these extra moments are not explained by small canting of the Fe spins of $\alpha = 1.3^\circ$ away from the easy $b$-axis~\cite{toft2015anomalous,yiu2017hybrid} as will be shown below by HF-ESR data.

In order to further analyse the observed quasi-free moments, we have separated the non linear from the linear part by fitting the high-field behaviour with a linear function and subtracted the resulting straight line from the data~\cite{klingeler2005weak}. This procedure gives $M_\mathrm{qf}=M-M_{\rm lin}$ as shown in Fig.~\ref{MB}. It shows the alignment of quasi-free moments towards their saturation value of $M_\mathrm{qf}^{sat} = 0.071(1)$~\mbfu\ at around 5.5~T. This is in-line with the $\chi$ vs. $T$ data in Fig.~\ref{MT}a which imply suppression of the steep Curie-like upturn at 7~T. The data presented in Figs.~\ref{MT} and \ref{MB} hence clearly evidence the presence of only weakly correlated anisotropic magnetic moments in \lfp . Comparing the non-linear contribution $M_\mathrm{qf}$ with a simple Brillouin function (see Fig.~\ref{MB}) indicates a more complex behaviour of the extra moment as compared to completely isolated spins $S=2$. This is particularly evident when the derivative $\partial M/\partial B$ is considered which -- in addition to general right-bending of $M_\mathrm{qf}$ -- shows two anomalies at $B^*_1 = 1.1$~T and $B^*_2 = 1.8$~T.

\subsection{High-frequency electron spin resonance}

\begin{figure}	
\includegraphics[width=1.0\columnwidth,clip] {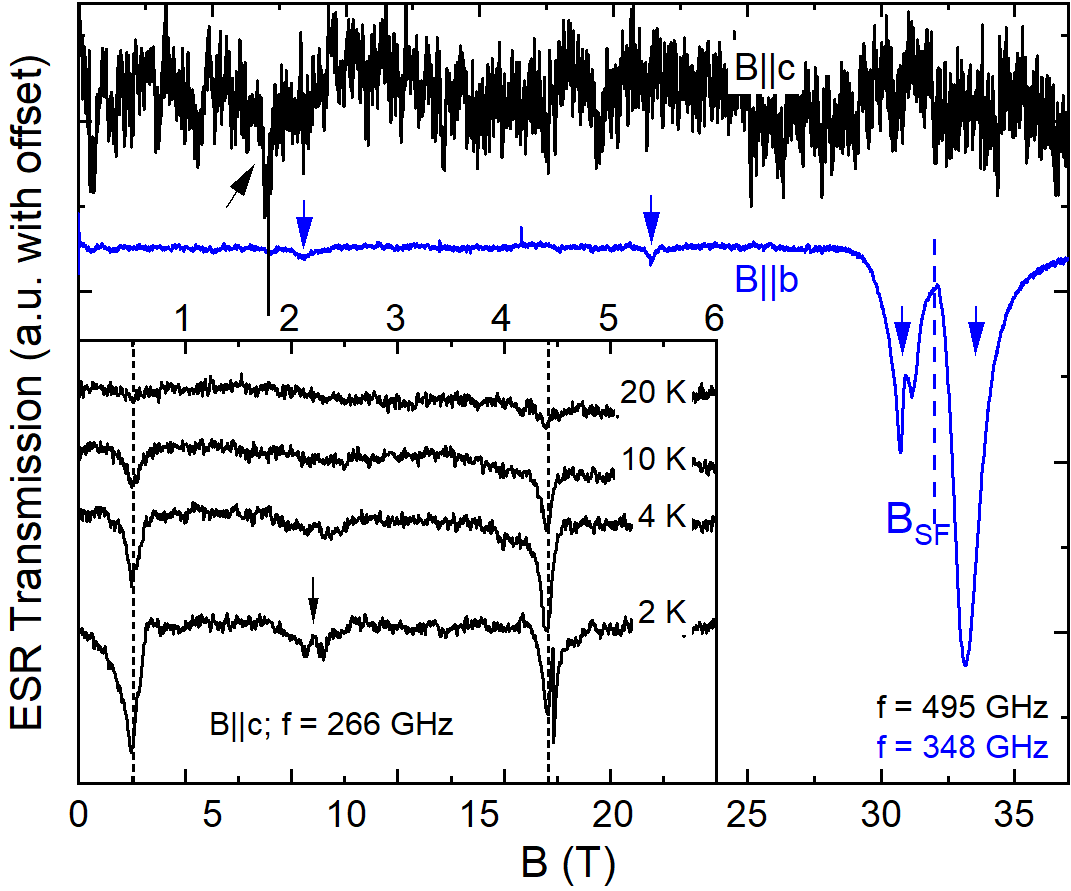}
\caption{High-frequency electron spin resonance spectra in pulsed fields $B||c$ and $B||b$ at $f = 495$~GHz and $f = 348$~GHz, respectively. The dashed line indicates the spin-reorientation field \bsf. Inset: ESR spectra with magnetic fields $B||c$ at $f = 266$~GHz. Dashed lines mark the resonance position at $T = 2$~K. Arrows indicate resonances.}\label{ESR}
\end{figure}

While HF-ESR is susceptible to collective magnetic excitations, at $q=0$, observation of magnon modes  in \lfp\ is challenging due to large single-ion anisotropy and strong exchange interactions which lead to large excitation gaps at zero magnetic field of 1446~GHz and 2072~GHz~\cite{li2006spin,toft2015anomalous,yiu2017hybrid}. When external magnetic fields are applied along the magnetic easy $b$-axis, collective resonance modes are however expected to soften in fields similar to the spin-reorientation field \bsf~\cite{nagamiya1955antiferromagnetism}. This is indeed observed in the HF-ESR spectra measured for $B||b$. The spectrum obtained at 348~GHz (Fig.~\ref{ESR}, blue line) features the two expected resonances close to \bsf = 32~T. In addition, two resonances with less intensity at lower fields are observed. Applying magnetic fields $B||c$, i.e., where below the zero field splitting no magnon modes are expected, yields a spectrum (black line) with at least one clear resonance of low intensity. One has to conclude that the weak features are not associated with magnon modes. Instead, the temperature evolution of the intensities of the weak resonance features implies a Curie-like behaviour. This is demonstrated by spectra taken at 266~GHz and for $B\|c$ displayed in the inset of Fig.~\ref{ESR}. The spectra display rather intense peaks at $\sim 0.5$~T and $\sim 4.4$~T as well as weaker slightly split resonances at $\sim 2.3$~T. All these features show Curie-like behaviour as the integrated intensities decrease upon heating, thereby resembling the Curie-like upturn of magnetic susceptibility (see Fig.~\ref{MT}a).

\begin{figure}	
\includegraphics[width=1.0\columnwidth,clip] {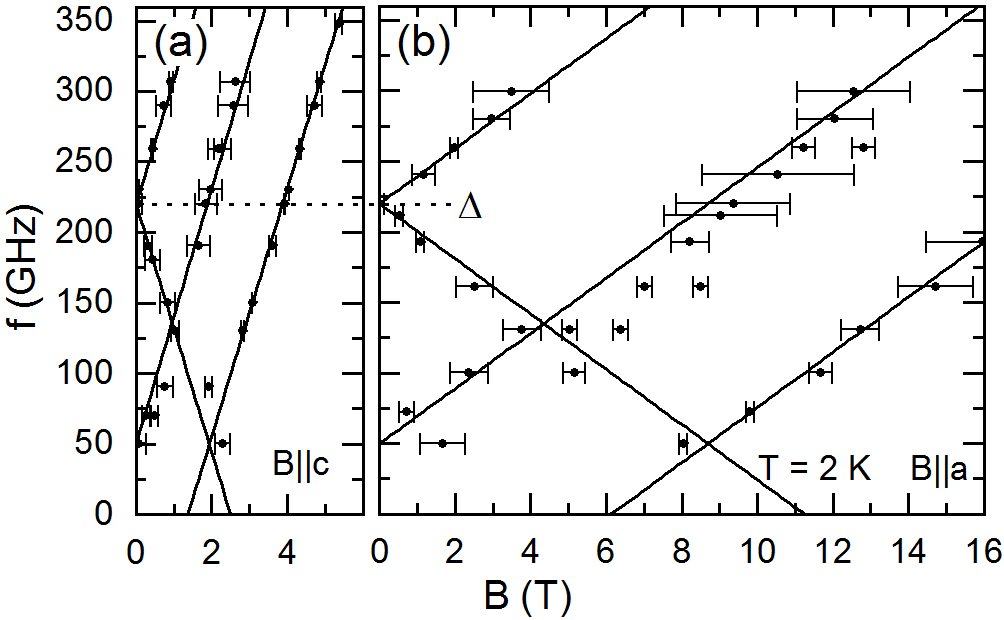}
\caption{Frequency dependence of resonance fields at 2~K with external magnetic fields (a) $B||c$ and (b) $B||a$. Solid lines are linear fits to the resonance branches. The dashed line indicates the zero field splitting $\Delta$.}\label{FB}
\end{figure}

The measured frequency dependencies of the resonance fields displayed in Fig.~\ref{FB} for magnetic fields $B||c$ and $B||a$, respectively, allows to clarify the microscopic origin of the these resonances. The general behaviour of the resonance branches is similar for both magnetic field directions. All resonances are associated with resonance branches with linear field dependence of resonance frequency and, except for the sign, similar slopes for a given field direction. In addition, for each field direction we observe two gapped modes of opposite slope with zero field splitting of $\Delta = 220(3)$~GHz. At approximately $\sim50$~GHz, the falling branches crosses or turn into an increasing mode. Such crossing of spin states is typically associated with anomalies in the static magnetisation and it indeed nicely accounts for the anomaly $B^*_2$ (see Fig.~\ref{MB}).
Finally, there is an additional mode with positive slope with zero field splitting of $\Delta'\sim50$~GHz.

The slopes of the branches allow determining the respective effective $g$-values which in the following will be just named $g$-values. As mentioned above, the slopes and hence the $g$-values of all branches of each direction are equal within error bars while there are large differences for the different field orientations. Quantitatively, the data in Fig.~\ref{FB} imply $g_c = 6.3(1)$ and $g_a = 1.4(1)$ \footnote{Applying magnetic fields along the $b$-axis results in a more complex frequency dependence of the resonance fields with a $g$-value of the main resonance lines of $g_b = 2.0(1)$}. The observation of a large $g$-value for $B||c$ and a rather small one for $B||a$ is fully consistent with the anisotropic Curie-like upturn of the susceptibility. The measured anisotropy $(g_a/g_c)^2$ would yield $C_\mathrm{a} = 0.009(2)$~erg K/G$^2$ mol.

\section{Discussion}

Both static and dynamic magnetic studies evidence the presence of anisotropic magnetic moments which are not involved in long-range antiferromagnetic order. As Fe moments in the ground state are slightly tilted, it is illustrative to compare the resonance branches presented above with modes originating from tilted moments which, e.g., are observed in $\alpha$-Fe$_2$O$_3$~\cite{pincus1960theory}, $\alpha$-Cu$_2$V$_2$O$_7$~\cite{wang2018magnetoelastic}, and Sr$_2$IrO$_4$~\cite{bahr2014low}. In all examples, low-energy excitation modes show bending, i.e., are clearly non-linear and exhibit stronger anisotropy of the effective $g$-values as observed in \lfp. In addition, the resonance modes of tilted moments are usually not observed for all magnetic field directions. This further confirms that extra moments in \lfp\ are not associated with the long-range ordered ground state of Fe$^{2+}$ moments on Fe lattice sites. Instead, defects which are strongly associated with the crystallographic lattice of \lfp\ are accountable for the observed anisotropic Curie-like response. We note that this conclusion is supported by the fact that the saturation value $M_\mathrm{qf}^{sat}(B\| c)$ of the quasi-free component of the magnetisation is sample dependent.

The obtained Curie-constant $C_\mathrm{c}$, the saturation magnetization of the extra moments $M_\mathrm{qf}^{sat}$, and the $g$-value $g_\mathrm{c}$ enable determining the effective total angular momentum quantum number
\begin{align}
J = \frac{C_\mathrm{c}}{M_\mathrm{qf}^{sat}} \frac{3 k_\mathrm{B}}{\mu_\mathrm{B}N_\mathrm{a}g_\mathrm{c}}-1
\end{align}
of the impurities. This procedure yields $J = 2.16(9)$ which is close to the integer $J_\mathrm{eff} = 2$ and hence suggests attributing the anisotropic moments to Fe$^2+$ impurities with high spin $S = 2$ and quenched orbital momentum $L \approx 0$. This leads to a density of impurities of $n = 0.0053(7)/\mathrm{f.u.}$.

Low-temperature increase of the susceptibility is not only observed for \lfp\ but was reported previously for other phosphates, too. In LiMnPO$_4$ it however shows isotropic Curie-like behaviour~\cite{neef2017high}. This discrepancy is attributed to an isotropic $g$-factor expected for half-filled d-shell of Mn$^{2+}$-ions. For LiMn$_{1-x}$Fe$_x$PO$_4$ it was observed, that with increasing iron content the Curie-like susceptibility upturn is getting more anisotropic~\cite{neef2017high}. An even stronger increase of the susceptibility in $c$-direction is reported for \lfp\ doped with non-magnetic Mg-ions~\cite{chen2007magnetic}. Our results suggest that this behaviour is directly linked to the increase of anti-site disorder.

The crystal structure exhibits four different Li-sites with the same local distorted octahedral oxygen environment. The distorted octahedra exhibit $C1$ symmetry and are positioned in the crystal structure in a way, that always the same facet of the octahedron is normal to the $c$-axis of the crystal structure. The octahedra are however rotated in the $ab$-plane. For Fe at a Li-position, the low-symmetry coordination hence fully lifts orbital degeneracy which results in orbital splitting into separated $m_\mathrm{s} = \pm 1$ and $m_\mathrm{s} = \pm2$ doublets and a singlet $m_\mathrm{s} = 0$. This general energy level schema is confirmed by the HF-ESR data as the observed low-field ESR branches are straightforwardly associated with allowed transitions between split spin states of localised $S=2$ (see Fig.~\ref{struc}). In particular, the modes with zero-field gap of $\Delta = 220(3)$~GHz display transitions between $m_\mathrm{s} = \pm 1$ and $m_\mathrm{s} = \pm2$ states and $\Delta$ quantifies their splitting in zero field. While, the resonance branch with $\Delta'\sim50$~GHz is associated with transitions between $m_\mathrm{s} = 0$ and $m_\mathrm{s} = \pm 1$ states. Consequently, anomalies $B^*_1$ and $B^*_2$ in the magnetisation curve signal the changes of the magnetic ground state from $m_\mathrm{s} = 0$ to $m_\mathrm{s} = -1$ and $m_\mathrm{s} = -1$ to $m_\mathrm{s} = -2$, respectively.

Comparing our experimental results to materials with paramagnetic Fe$^{2+}$ ions in a low-symmetry coordination, the here observed zero-field splitting of the energy levels is in the common rage \cite{krzystek2006multi,rudowicz2003characteristics}. In contrast, the effective $g$-value respectively the $g$-value anisotropy observed in our study is unusually high as, e.g., in Mg$_2$SiO$_4$:Fe where Fe$^{2+}$-ions are on octahedral sites of the forsterite structure, anisotropic $g$-values of $g_{||} = 4.30$ and $g_\perp =2.0$ are observed \cite{shakurov2011high}.

Since formation of defects depends on details of the synthesis process, Fe$^{2+}$-antisite density is supposed to vary between different samples and different parts of the crystalline rod obtained by our optical floating-zone method. Indeed, while $n=0.53(7)$\% is obtained from the analysis of magnetic data above, a different part of the same single crystalline batch which exhibits identical \tn\ shows $n=0.32$\% (see the the supplement). Both values are in perfect accordance with the expected equilibrium defect concentration of 0.1 to 0.5\% for the used solid-state synthesis temperatures \cite{malik2010particle}.

In comparison with other experimental methods, our results quantitatively agree to determination of antisite disorder by lithium diffusion experiments on crystalline nanoplatelets of \lfp , where 0.5\% antisite concentration was obtained \cite{liu2017effects}. Diffraction studies illustrate the large regime of potential antisite defect concentrations in \lfp . High concentrations can be, e.g., due to insufficient solid state diffusion so that values of 7 to 8\% have been reported for materials from hydrothermal and high-temperature solid state synthesis methods~\cite{yang2002reactivity}. The growth method of single crystals implies rather diverse antisite concentrations as flux-grown \lfp\ is reported to exhibit no evidence for antisite defects in these samples~\cite{janssen2013reciprocal} from a combination of powder and single crystal diffraction (X-ray and neutron). While, synchrotron powder X-ray diffraction data on ground crystals grown by the optical floating-zone (TSFZ) method at ambient pressure imply values of 2.5\%~\cite{amin2008ionic}. Laboratory powder XRD data on single crystals of the same batch as studied here have been best refined when antisite concentration of $n = 0.023(2)/\mathrm{f.u.}$~\cite{neef2017high} was considered which is larger than the concentration of quasi-free $S=2$. We attribute this to the uncertainty of the powder XRD method with respect to a particular kind of defects.~\cite{janssen2013reciprocal} We conclude, that in addition to the particular paramagnetic-like type of defects associated with Fe$^{2+}$ on Li-sites further defects-types may be included in the values provided by Rietveld analysis of conventional x-ray patterns. We emphasize in this respect, that our magnetic studies can discriminate between different defects. While non-magnetic ones are not detected, different magnetic defects would result in different $g$-factors and zero-field splittings which is not observed. We mention that failure of detecting Li-Fe antisite defects by means of X-band ESR in Ref.~\onlinecite{amin2006anisotropy} is fully in-line with the fact that zero-field splitting is larger than 10~GHz (see Fig.~\ref{FB}) which prevents detection in conventional X-band ESR.

In addition, our results imply that in theoretical studies on electronic properties and antisite-defect mediated diffusion in \lfp\ orbital degrees of freedom should be taken into account, which can be concluded from the highly anisotropic $g$-factor of the defect moments. In particular, a local structure model as used in Ref.~\onlinecite{adams2010lithium} can not account for these features. In general, DFT calculations of antisite defects~\cite{dathar2011calculations} should yield the anisotropic $g$-values as well as zero-field splitting due to crystal field effects which may be used to confirm the validity of a numerical approach.

Knowledge of the strongly anisotropic $g$-values also enables to quantitatively determine antisite disorder by static magnetometry even in \lfp\ powders. A Curie-Weiss fit to the volume magnetic susceptibility in the temperature region of 2~K to 20~K results in a powder averaged Curie constant $C$ from which the defect concentration

\begin{align}
n = \frac{3C\, k_\mathrm{B}}{2\mu_\mathrm{B}^2N_\mathrm{A}(g_\mathrm{a}^2+g_\mathrm{b}^2+g_\mathrm{c}^2)}.
\end{align}

is obtained by using $g_\mathrm{a} = 1.4$, $g_\mathrm{a} = 2.0$ and $g_\mathrm{c} = 6.3$.

\subsection{Summary}

Our detailed studies of the static and dynamic magnetic properties on single crystals by means of magnetisation and high-frequency electron spin resonance measurements enable qualitative and quantitative conclusions on the nature and the properties of antisite defects in \lfp . The presence of rather localised moments which are only weakly magnetically interacting with the magnetic subsystem on the crystallographic Fe-positions is attributed with Fe$^{2+}$-ions on Li-positions. The fingerprint of these moments includes linear resonance branches, a highly anisotropic $g$-factor with $g_\mathrm{a} = 1.4$, $g_\mathrm{b} = 2.0$, and $g_\mathrm{c} = 6.3$, and significant zero-field splittings of $\Delta = 220(3)$~GHz and $\Delta' \sim 50$~GHz. We demonstrate a procedure to precisely determine the defect concentration by static magnetisation measurements on powder materials.

\begin{acknowledgements}
The project is supported by Deutsche Forschungsgemeinschaft (DFG) through KL 1824/13-1. We acknowledge the support of the HLD at HZDR, member of the European Magnetic Field Laboratory (EMFL).
\end{acknowledgements}

\end{document}